# Temperature of the SN1987A Supernova Provides an Estimate of the Electron Neutrino Mass*

by John Michael Williams

P. O. Box 2697,
Redwood City, CA 94064
jwill@AstraGate.net

# Abstract

Numerous upper bounds on the (anti)neutrino rest mass have been published based on the SN1987A observations.  Here, we use a nonkinematic (thermal) time extent to provide a rest-mass estimate of a few eV (as $mc^2$ energy), if not zero.  In the solution, we find that a typical upper-bound formula for the mass implies that this thermal extent was attributable to about 10% of the particle energy measured on Earth.  The present approach yields an expected value for the mass, given any theoretical or model-dependent estimate of the fraction of the detected neutrino energy attributable to the supernova temperature.

# I.  Introduction

The theoretical proposal of the neutrino particle in 1930, by Pauli, was followed by the experimental demonstration of the existence of the electron neutrino in 1953 by Reines.  At the time, it was assumed that, like the photon, the electron neutrino likely would have a rest mass of zero [1].  The current Standard Model of elementary particles postulates massless neutrinos [2]; however, many theorists in astrophysics and particle physics believe that all the leptons, including the neutrinos, should be massive [32; 3-5].  At 5 $\text{eV}/c^2$, the rest mass would be about $10^{-5}$ that of an electron; this doesn't seem unreasonable, although it certainly is not proven either by the current paper or any previous work.

The temporal interval of an observed particle stream from a distant event may be analyzed into three components:  (1) The *creation interval* of the primitive event, (2) the dispersion interval of the particle creation kinematics (here called the *energy extent*), and (3) the dispersion interval because of the temperature during creation (here called the *thermal extent*).  The total time interval at the detector is the *detection interval*.  In the case of supernova neutrinos, the center of mass motion of the primitive region, assumed very subluminal relative to Earth, would not affect any of the time intervals measurably; and, no supernova process would be expected to reduce the times, under the First and Second Laws of thermodynamics, unless by rare chance, a possibility we here ignore.

If the electron (anti)neutrino had zero mass, the transit from a supernova to Earth would be assumed to have been at the exact speed of light in a vacuum, making the detection interval always equal to the creation interval and the other components zero; however, a neutrino with nonzero rest mass may not be allowed the speed of light.

## II.  The 1987 Supernova Burst

The Kamiokande II and Irvine-Michigan-Brookhaven (IMB) neutrino detectors were designed to study proton decay and Solar neutrino flux.   Both were operating in February of 1987, when the SN1987A supernova event occurred in the Greater Magellanic Cloud at a distance of about 50,000 parsecs.   The star causing this supernova has been identified and was catalogued as a blue giant of about 15 Solar masses before the supernova [6].   The Kamiokande II data, with inferred neutrino energies accurate to about 20% [7], are plotted in Fig. 1; the IMB data were similar but show fewer neutrinos over a shorter time interval.

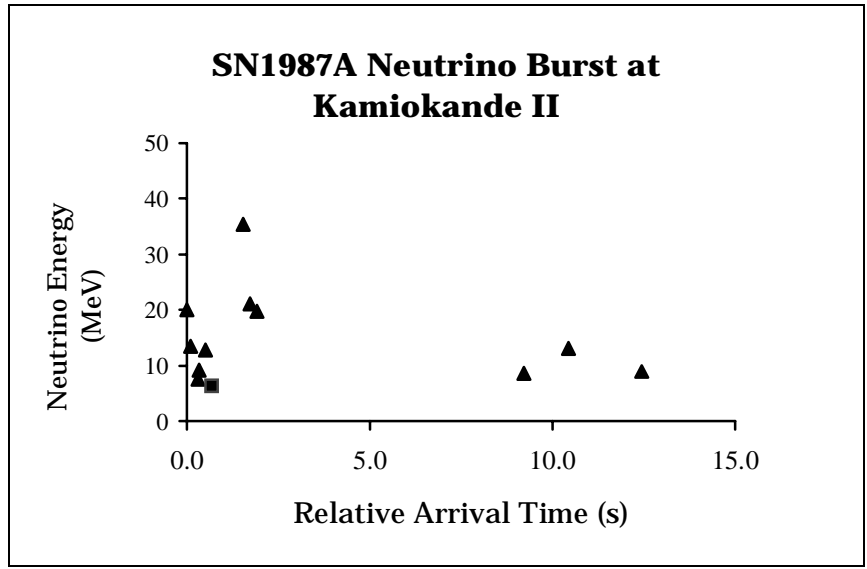

**FIG. 1. Antineutrino energy at the detector, assuming the Cerenkov light represented antineutrino+proton → positron+neutron interactions. Data after Hirata et al [8, Table 1]. The square symbol represents the one likely detector background event.**

Looking at Fig. 1, the maximum detection interval possible would seem to be about 15 seconds. The period up to about 2 seconds also might be considered the detection interval for a distinct subevent. Supernova collapse and opacity models have suggested a primary neutrino emission process which would decay exponentially with a time constant on the order of 3 seconds [33], based in part on these data.

# III.  Thermal Extent yields an Estimate of the Mass

## A.  Neutrino thermal velocity component

Let us assume the neutrino to have nonzero rest mass.   Also, let us assume (*a*) that a gas of boron $^8$B atoms (*b*) at approximately $10^9$ K [9] was the primitive substance for the neutrinos detected.

Consider the effect of an error in assumption (*a*):  Primitives, say, atoms or stripped nuclei, heavier than $^8$B boron, such as $^{28}$Si or $^{56}$Fe, would reduce the thermal velocity estimate (to be given below) by no more than a factor of three; for all else the same, this would require slightly more massive neutrinos to account for the observed detector energies.   But, because the creation interval would include any subsequent weak-force diffusion period, the kinetic temperature probably should reflect a primitive lighter than these heavy-element fusion products.   So, we will accept $^8$B as the primitive substance somewhat arbitrarily and without further consideration of error.    We accept assumption (*b*) as reasonable and will comment on a range of temperatures later.

The thermal extent by definition is meant to ignore all kinematic energy differences [10] and to sidestep complexities [4, 9, 11-13] of state during the creation interval; these factors will be seen later to be described in the energy, not thermal, extent of the observations.

The observed neutrinos being assumed massive, the primitive thermal motion would add a random component to the velocity of the neutrino as measured in the inertial frame of the detector on Earth. We estimate this velocity dispersion by equating the primitive kinetic and thermal energy as

$$\frac{1}{2} m_p \langle v_p \rangle^2 = \frac{3}{2} k_B T_p, \tag{1}$$

in which $m_p$ and $\langle v_p \rangle$ respectively represent the rest mass and rms average thermal speed of a primitive element during creation of the detected neutrino, $k_B$ is Boltzmann's constant, and $T_p$ is the Kelvin temperature in the primitive region of the supernova.

Solving for $\langle v_p \rangle$, which we shall use to define the average inertial frame of the detected neutrino particles, we get

$$\langle v_p \rangle = \sqrt{3 \frac{k_B}{m_p} T_p} \; . \tag{2}$$

The rest mass of $^8$B would be a little over 8 times that of the proton, making $m_p$ about $1.5 \cdot 10^{-26}$ kg. Assuming $T_p$ to be somewhat over $10^9$ K, a rough estimate of the speed from eq. (2) would be $\langle v_p \rangle \cong 3 \cdot 10^6$ m/s $\cong 0.01c$.

For simplicity, we accept this average speed of about $0.01c$ as mapping directly to the thermal extent. We shall double it to convert it from an absolute-value scalar to the one-dimensional magnitude of the temperature factor in a neutrino velocity vector on the line of flight between the supernova and Earth. The thermal extent may be scaled, if desired, based on the reader's favorite probability theory (e. g., [14]); the creation spectral distribution has been described as Planckian [15]. We note that for purely thermal particles, we would add the creation interval, left for now as an unknown, to the thermal extent to get the detection interval.

So, taking the distance from the <u>S</u>upernova to <u>E</u>arth as $s_{SE}$, the spatial line-of-flight dispersion of interneutrino distance at Earth as $ds_{SE}$, and the speed of the neutrino as $v_{SE}$, the <u>t</u>hermal extent $dt^t_{SE}$ may be computed from the supernova (or Earth) proper transit time $t_{SE}$ using these relations:

$$t_\nu = \frac{t_{SE}}{\gamma} = \frac{s_{SE}}{\gamma v_{SE}}, \qquad \text{with } \gamma = \left(\sqrt{1-\left(\frac{v_{SE}}{c}\right)^2}\right)^{-1} ; \text{ and,} \qquad (3)$$

$$dt^t_{SE} = 2\frac{ds_{SE}}{\gamma v_{SE}} = 2\frac{\langle v_p \rangle t_\nu}{\gamma v_{SE}}, \qquad (4)$$

in which $t_\nu$ is the neutrino-proper transit time and $\gamma$ gives the Lorentz (special relativity) transformation. As already mentioned, the factor of 2 in eq. (4) is because of the assumed bidirectional random projection of the thermal velocity of any given neutrino on the direction vector of the supernova-Earth line of flight.

Defining the conversion factor $psk = (3.26)(60^2)(24)(365)c = 3 \cdot 10^{16}$ for meters per parsec, eqs. (3) and (4) may be combined into one formula for thermal extent,

$$dt^t_{SE} = 2\frac{\langle v_p \rangle s_{SE} psk}{v_{SE}^2}\left[1-\left(\frac{v_{SE}}{c}\right)^2\right]. \qquad (5)$$

## B. Rest mass from velocity and thermal extent

In general, the total energy $E$ of any free particle with momentum $p$ and rest mass $m$ in transit in a vacuum is given by

$$E^2 = (pc)^2 + (mc^2)^2 \tag{6}$$

Solving (6) for *m* after substituting $mv\gamma$ for *p* yields, for the (anti)neutrino,

$$m_v = \frac{E\sqrt{c^2 - v^2}}{c^3} = \frac{E\sqrt{2c(c-v) - (c-v)^2}}{c^3}. \tag{7}$$

Now, (7), (5), and (2) yield the rest mass formula, depending on energy $\langle E^t \rangle$ of the thermal extent,

$$m_v^t = \frac{\langle E^t \rangle}{c} \sqrt{\frac{dt_{SE}^t}{c^2 dt_{SE}^t + 2(s_{SE} psk)\sqrt{3\frac{k_B}{m_p}T_p}}}, \tag{8}$$

the superscript indicating the energy by assumption exclusively contributing to the thermal extent.

We note that the thermal speed $\langle v_p \rangle$ has been eliminated in favor of temperature and thermal energy, and that there is no assumption of any spread in particle energies in (8). Also, (8) is an expected-value estimate, depending solely on the bounds of its parameters; it is not a formula for an upper or lower bound.

## IV. Energy Extent yields an Upper Bound on the Mass

A proposed analysis based on energy dispersion was published by Zatsepin [16] in 1968, long before data were available; he suggested that supernova data could be used to estimate an upper bound on the neutrino rest mass at least down to about 2 $eV/c^2$. The subsequent literature [10, 12-15, 17-31] based on energy analysis of the

J. M. Williams                     Neutrino Temp v. 2..0                     9

SN1987A data have concluded neutrino mass upper bounds in the low tens of eV or less. Anada and Nishimura [27] used a temperature-based coherence argument to eliminate the SN1987A data as a way of evaluating the theory of neutrino type oscillations. Roos [24] mentioned the possibility of calculating a thermal-dispersion bound on the neutrino rest mass but evidently did not attempt a formula for it. Abbott et al [21] used both energy and thermal parameters in a differential formulation which seemed to yield a zero rest mass because zero fell near the middle of a fairly broad error range. At least one author [19] has asserted that "model-independent" limits are not reliable. The majority of the published papers on the subject introduce specific supernova or particle assumptions.

Reasonable model-independent formulae for the energy extent usually assume a neutrino creation interval of a few seconds or less. Here, we adopt a variation of a formula in [10; *cf.* 24] as follows,

$$dt_{SE}^{E} = \frac{s_{SE}}{c}\left(\frac{m_\nu c^2}{E^E}\right)^2, \qquad (9)$$

in which $dt_{SE}^{E}$ is the <u>e</u>nergy extent, $s_{SE}$ is as above, and $E^E$ is the energy of the detected neutrino. The detected energy of course would include any thermal kinetic component determined by the primitive temperature. There is an implied assumption that the creation interval amounted to a negligible time, so that the energy extent would refer to all or most of the detection interval; namely, $\Delta E^E / E^E \cong 1 \Rightarrow duration(\Delta E^E)/duration(E^E) \cong 1$. This yields an upper bound on the mass: The $\Delta E^E$ cannot reasonably exceed $E^E$; and this formula ignores the kinetic term in (6) above, moving supernumerary energy into the $mc^2$ mass.  Solving (9) for the neutrino mass $m_\nu$ yields

$$m_\nu^E = \frac{E^E}{c}\sqrt{\frac{dt_{SE}^E}{c(s_{SE}\,psk)}}\,, \tag{10}$$

which may be compared with the formula (8) above for $m_\nu^t$.

## V.  Implications of the Two Rest Mass Formulae

Suppose the neutrino creation kinematic energy was constant, so all neutrinos were created with the same energy, but with different thermal variation.  Then, the thermal extent would equal the energy extent and not merely be included in the latter.  In this case, (8) might be used alone to estimate the rest mass.

We here accept that the kinematic energy was not equal for every neutrino, and so might be used to account for any assumed difference between creation and detection intervals.  The thermal (8) and energy (10) estimates each yield a mass based on an unknown overlap in the thermal *vs.* energy contributions.  However, in neither formula is the overlap defined; thus, in neither is it constrained.  So, having assuming a nonzero rest mass, we equate the $m_\nu^t$ and $m_\nu^E$ masses (8) and (10) of the neutrino and arrive at the following formula,

$$dt_{create} = dt_{SE} - \frac{s_{SE}\,psk}{c}\left[\left(\frac{E^t}{E^E}\right)^2 - \frac{2}{c}\sqrt{3k_B\frac{T_p}{m_p}}\right], \tag{11}$$

in which the expected-value brackets for $E^t$ have been dropped.  Using the trial values, $dt_{SE} = 10\text{s}$ and $E^E = 10$ MeV, we find that $dt_{create}$ from (11) can be within a few seconds of zero only for values of $E^t$ about as shown in Table 1. $E^E$ only implies an upper bound on the mass because of (10); it is independent otherwise of any bound. So, the value of $E^E$ might be used directly in place of $E^t$ in (8) for a mass estimate, provided a model or theory could account for the thermal fraction at the detector.

Table 1 shows that 10 MeV neutrinos created at a primitive temperature of $10^9$ K would require $E^t$ near 1 MeV. Happily, the range of the energy ratio near $dt_{create} = 0$ in (11) is relatively insensitive to the value of $dt_{SE}$ chosen; the Table 1 dependence on primitive temperature is relatively mild in the region tabulated.

**TABLE 1. Neutrino energy in the thermal extent, for $E^E$ = 10 MeV, yielding a creation interval near zero in eq. (11), for various supernova temperatures.**

| Primitive Temperature $T_P$ (K) | Energy of Thermal Extent $E^t$ (MeV) |
|---|---|
| $10^8$ | 0.611 |
| $10^9$ | 1.08 |
| $10^{10}$ | 1.93 |
| $10^{11}$ | 3.43 |
| $10^{12}$ | 6.09 |

Finally, Fig. 2 displays the rest mass here estimated for the first time by the thermal extent; it also shows how the energy extent provides an upper bound as ordinarily estimated. Taking the detection interval as 2 s, which seems not unreasonable from the Kamiokande II data in Figure 1, we conclude that the mass would be a few $eV/c^2$ and can not exceed about 6 $eV/c^2$. If we were allowed the additional assumption that the creation interval was, say 1.5 s, we could use Figure 2 to conclude a neutrino mass of about 3 $eV/c^2$. Conversely, if we knew the neutrino mass to be, say, 1 eV, we would know that the creation interval was just 1.8 s. As shown by the differences between the thermal-extent and energy-extent-bound curves in Figure 2, as the assumed creation interval decreases, the thermal extent comes to dominate the detection interval, regardless of the supernova model. The same dominance would be expected under an increase in the assumed supernova primitive temperature.

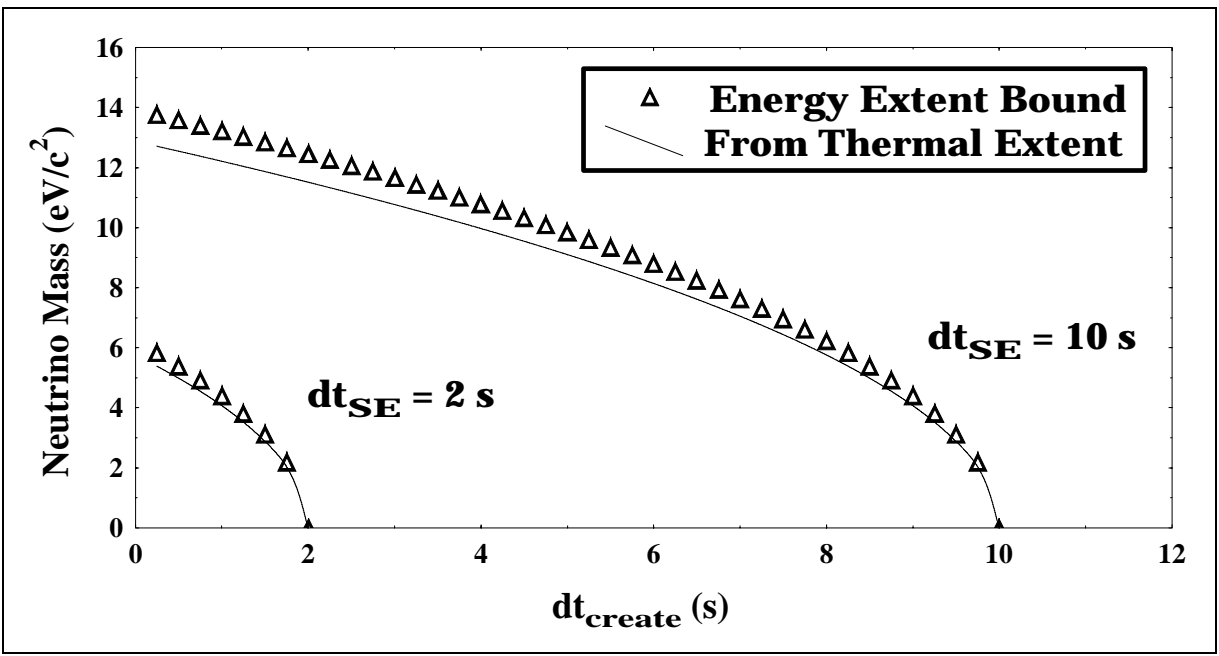

**FIG. 2. Rest mass of the SN1987A neutrinos as a function of the creation interval, from eqs. (8) and (10), for two values of the detection interval, $dt_{SE}$.**

# VI.  Conclusion

We have shown that the thermal line-of-flight dispersion of the neutrinos from supernova SN1987A might be used to estimate the rest mass, knowing the fraction of neutrino energy at the detector attributable to temperature.   Then, assuming massive neutrinos created in an interval not exceeding 10 seconds and at a boron-like primitive temperature of $10^9$ K, we used an upper-bound formula to find that the thermal fraction might reasonably account for about 10% of the neutrino energy at Kamiokande II.   From this, not the energy upper bound *per se*, it followed that the neutrino rest mass, here assumed not zero, would be as much as a few $\mathrm{eV}/c^2$ .

The present approach assumed nothing about the supernova dynamics but still yielded an expected value, not a bound.   Any supernova model providing a better estimate of the thermal energy fraction would yield a better expected value.

# Acknowledgements

The author thanks Kai Martens for a copy of the Kamiokande II data for the present analysis.   Thanks also are due to Mario Rabinowitz, for reading the paper and suggesting some improvements.

This paper was not supported by any grant to the author, who is an independent investigator, but it was inspired by presentations attended at the SLAC Summer Institute, *The Physics of Leptons*, given in August 1997.